\documentclass[aps,twocolumn,prb]{revtex4}  

\usepackage{graphicx}   		
\usepackage{amssymb}            	
\usepackage{url}			

\newcounter{tempref}

\begin{document}

\title{Resource Letter Exo-1: Exoplanets}
\author{Michael\ Perryman}
\affiliation{Department of Astrophysical Sciences, Princeton University}  
\email{mac.perryman@gmail.com}

\date{\today}

\begin{abstract}
This Resource Letter gives an introduction to the main topics in exoplanet research.  It is intended to serve as a guide to the field for upper-division undergraduate and graduate students, both theoretical and experimental, and for workers in other fields of physics and astronomy who wish learn about this new discipline. Topics include historical background, detection methods, host star properties, theories of planet formation and evolution, their interiors and atmospheres, their relationship to the formation and evolution of our own solar system, and issues of life and habitability. 
\end{abstract}

\maketitle

\section{Introduction}

Throughout recorded history, mankind has speculated about the existence of other planets, and the possibility of life, beyond our own solar system. The unambiguous detection of the first such extra-solar planets, or `exoplanets', in the early 1990s has transformed the field into a rapidly expanding and quantitative field of astronomical research. A variety of observational methods have been developed for their discovery and characterisation.

As of 1~November 2013, just over 1000 exoplanets (in nearly 200 multiple systems) are considered as `confirmed', many through dedicated space observations (and many more expected from the ongoing Kepler satellite analysis). The architecture of many of the systems (in terms of masses and orbits) are qualitatively different than those that had been expected on the basis of our own solar system. Explaining their properties has led to a substantial advance in understanding their formation, evolution, and internal structure, as well as the formation and evolution of our own solar system.

The literature accompanying their discovery and characterisation is substantial. Since 1990, some 7000 papers have been published in the refereed literature (and an estimated 3000 or more in conference proceedings), rising from 50--100 per year until 2000, 300--400 until 2005, 500 around 2010, and exceeding more than 1000~per year over the past 2--3~years. These cover discoveries and physical properties (and related instrumental, observational, and data analysis approaches), as well as theoretical, modeling and other interpretative aspects including structure, formation, evolution, composition, and `habitability'. 

Evidently, given this extensive literature, the following can only represent a selection, chosen to provide some orientation within this multi-disciplinary field.

\begin{figure*}[t]
\centering
\frame{\includegraphics[width=1.0\linewidth]{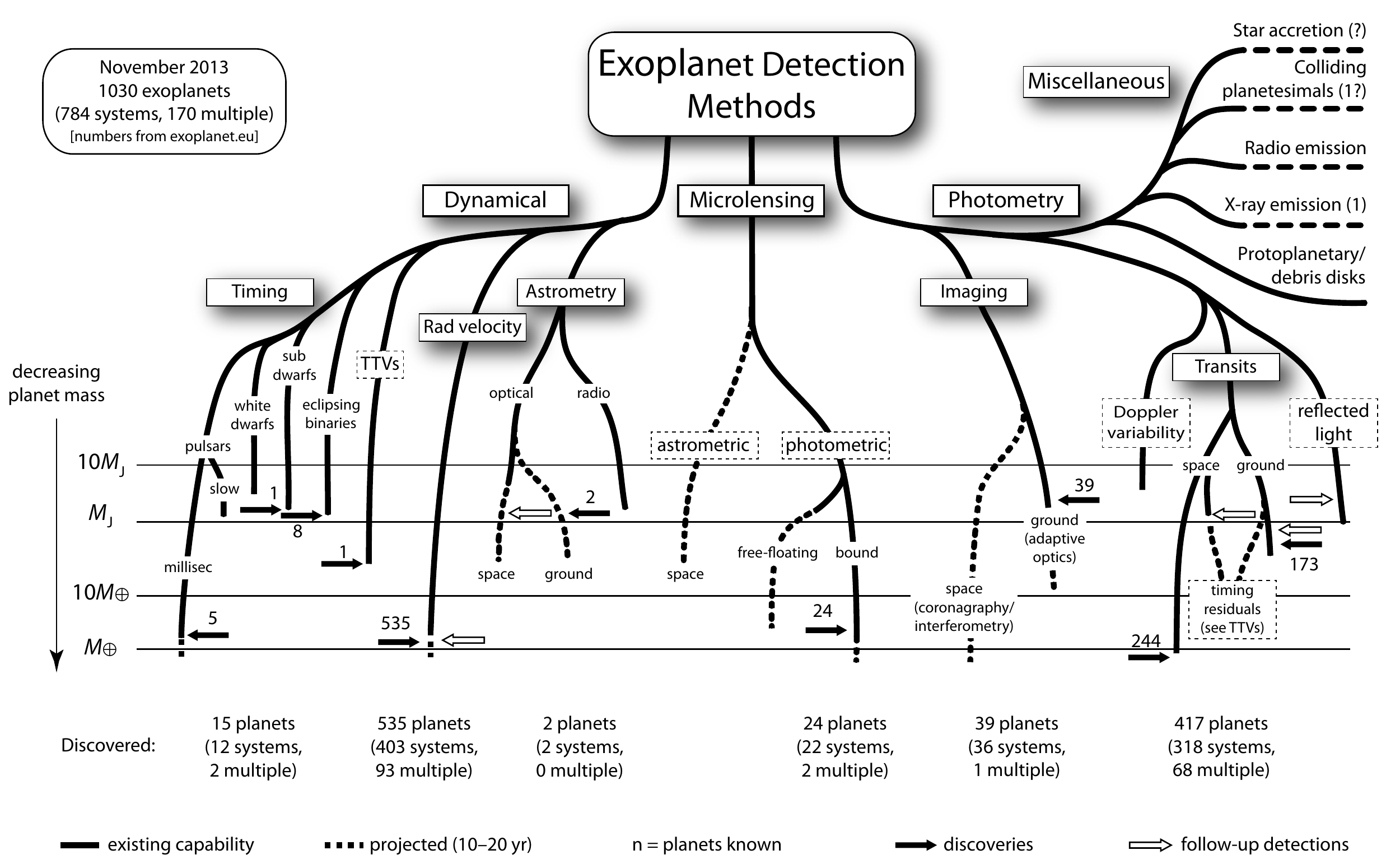}}\\[-2pt]
\caption{Detection methods for exoplanets. The lower limits of the lines indicate masses that are detectable by present measurements (solid lines), and those that might be expected within the next 10~years (dashed). The (logarithmic) mass scale is shown at left. The miscellaneous signatures to the upper right are less well quantified in mass terms. Solid arrows indicate detections according to approximate mass. Open arrows indicate that relevant measurements of previously-detected systems have been made. The figure takes no account of the numbers of planets that may ultimately be detectable by each method.
\label{fig:detection-methods}
}
\end{figure*}

\section{Basic resources}
\label{sec:basic-resources}

\noindent
{\it Journals:} some 10~journals publish the majority of refereed research papers on exoplanets (with the approximate number published between 1995--2012 given in parentheses):
as for other scientific fields, {\it Nature} (200) and {\it Science} (80) feature {\it some\/} of the more prominent advances, mostly in the observational domain. The other main journals are
{\it The Astrophysical Journal} (ApJ, 2500),
{\it Astronomy \& Astrophysics} (A\&A, 1280)
{\it Monthly Notices of the Royal Astronomical Society} (MNRAS, 900),
{\it The Astronomical Journal} (AJ, 280),
{\it Publications of the Astronomical Society of the Pacific} (PASP, 200),
and
{\it Publications of the Astronomical Society of Japan} (PASJ, 200).

Articles connected to solar system aspects also appear in {\it Icarus} (220)
while those related to astrobiology and habitability also appear in {\it Astrobiology\/} or the {\it International Journal of Astrobiology\/} (200~together).
Numerous instrumental developments are found in {\it Proc.\ Society of Photo-Optical Instrumentation Engineers} (SPIE).

Synopses appear in various conference proceedings, with more formal reviews appearing in {\it Annual Reviews of Astronomy \& Astrophysics} (ARAA) and {\it Astronomy \& Astrophysics Reviews} (A\&ARv).

\begin{enumerate}
\itemsep -3pt
\footnotesize
\setcounter{enumi}{\thetempref}

\item 
{\bf ADS}
The SAO--NASA Astrophysics Data System (\url{adsabs.harvard.edu/abstract_service.html}) is the {\it de facto} digital library portal for publications in physics and astronomy. The ADS bibliographic database contains more than 10~million records, provides abstracts, links to the journal, and full-text scans, searchable through flexible queries.
 
\item
{\bf arXiv preprints} Many articles in astronomy appear first (although not necessarily in peer-reviewed form) in an online archive maintained by Cornell University at \url{arXiv.org} (and linked to the ADS), with those of relevance to exoplanet research appearing in the `physics/astro-ph' section.

\setcounter{tempref}{\theenumi}
\end{enumerate} 

\vspace{10pt}\noindent
{\it Textbooks and Monographs:} Complementing the more specialized texts referred to below, few deal with the topic in its entirety.

\begin{enumerate}
\itemsep -3pt
\footnotesize
\setcounter{enumi}{\thetempref}

\item \label{item:perryman}
{\bf The Exoplanet Handbook},
M.A.C.~Perryman (Cambridge University Press, Cambridge, 2011).
Textbook covering all scientific aspects, including detection, formation, and composition. Comprehensive (some 3000) literature references. Figures also available via the CUP www site.~(I)

\setcounter{tempref}{\theenumi}
\end{enumerate} 

\vspace{10pt}\noindent
{\it Conferences and workshops:}
There are frequent meetings on a variety of associated topics. An up-to-date calendar is given at \url{exoplanet.eu/meetings}.

\vspace{10pt}\noindent
{\it www resources:}
Table~\ref{tab:introduction-onlineresources} lists some online catalogues and associated data of particular relevance. Of these, the Extrasolar Planets Encyclopedia, \url{exoplanet.eu}, maintains an up-to-date catalogue and bibliography of confirmed exoplanets detected by all methods, and a separate list of unconfirmed or retracted planets.

\section{Preamble}

\subsection{Historical speculation and first detections}

The reliable detection of the first exoplanets in the 1990s followed centuries of philosophical speculation about the existence of other worlds, various erroneous claims largely based on astrometric measurements made over earlier decades, on studies of the various observational prospects, and on radial velocity measurement programs aimed at characterizing the substellar binary mass distribution.

\begin{enumerate}
\itemsep -3pt
\footnotesize
\setcounter{enumi}{\thetempref}
\item
{\bf The Extraterrestrial Life Debate 1750--1900. The Idea of a Plurality of Worlds from Kant to Lowell}, M. J. Crowe (Cambridge University Press, Cambridge, 1986). A detailed scholarly study, with some 600 references, of the lengthy speculations on whether or not life exists on any other body in the universe.~(E)

\item
``The first high-precision radial velocity search for extra-solar planets,'' G.A.H~Walker, New Astronomy Reviews {\bf 56}, 9--15 (2012). Perspective on the first radial velocity detections by one of the instrument pioneers. As quoted by the author {\it `It is quite hard nowadays to realise the atmosphere of skepticism and indifference in the 1980s to proposed searches for extra-solar planets. Some people felt that such an undertaking was not even a legitimate part of astronomy.'}~(E)

\item
``The discovery of 51~Peg at Haute-Provence Observatory: the quest for precise radial velocity,'' M.~Mayor, in {\bf Tenth Anniversary of 51~Peg\,b}, eds L.~Arnold, F.~Bouchy, C.~Moutou (Frontier Group, Paris, 2006), pp. 1--9. Traces the development of radial velocity instruments, leading to the first confirmed detections of a planet around a main-sequence star, by the first author of the discovery paper.~(E)

\item
``The history of exoplanet detection,'' M.A.C.~Perryman, Astrobiology {\bf 12}, 928--939 (2012). Summarizes the different methodologies, erroneous claims, and first detections.~(E)

\setcounter{tempref}{\theenumi}
\end{enumerate} 

\begin{table*}[t]
\caption{Online catalogues and associated data of particular relevance to exoplanet research. 
\label{tab:introduction-onlineresources}
}
\vspace{-10pt}
\centering
\footnotesize
\begin{tabular*}{\textwidth}[t]{@{\extracolsep\fill} lll}	
\noalign{\vspace{5pt}}
\hline
\noalign{\vspace{2pt}}
URL&								Content& 						Comment\\
\noalign{\vspace{2pt}}
\hline
\noalign{\vspace{2pt}}
Exoplanet catalogues:\\
\quad \url{exoplanet.eu}&					Extrasolar planets encyclopedia&	Catalogue and bibliography (all methods)\\[-1pt]
\quad \url{exoplanets.org}&				Exoplanet data explorer&			Critical compilation of exoplanet orbits \\[-1pt] 
\quad \url{exoplanetarchive.ipac.caltech.edu}&	NASA exoplanet database&		Exoplanet archive, transit data, and statistics \\[-1pt]
\noalign{\vspace{2pt}}
Transit data:\\
\quad \url{idoc-corot.ias.u-psud.fr}&			CoRoT mission&				Mission data archive (includes MOST mission)\\[-1pt]
\quad \url{kepler.nasa.gov}&				Kepler mission&				Mission description and planetary candidates\\[-1pt]
\quad \url{wasp.le.ac.uk}&				SuperWASP project&			Includes discoveries and data archive\\[-1pt]
\quad \url{var.astro.cz/etd}&				Exoplanet transit database&		Compilation of all exoplanet transits\\[-1pt]
\quad \url{astro.keele.ac.uk/jkt/tepcat}&		Transiting planet properties&		Physical properties (planets \& brown dwarfs)\\[-1pt]
\quad \url{physics.mcmaster.ca/~rheller}&	Rossiter--McLaughlin effect&		Compilation of spin--orbit alignment measures\\[-1pt] 
\noalign{\vspace{2pt}}
Related catalogues:\\
\quad \url{dwarfarchives.org}&				M, L, T dwarfs&					Photometry/spectroscopy/astrometry\\[-1pt]
\quad \url{circumstellardisks.org}&			Circumstellar disks&				Catalogue of pre-main sequence/debris disks\\[-1pt]
\noalign{\vspace{2pt}}
General resources:\\
\quad \url{adsabs.harvard.edu}&			SAO--NASA ADS&				Bibliographic data portal, including arXiv\\[-1pt]
\quad \url{simbad.u-strasbg.fr}&			CDS SIMBAD&					Comprehensive astronomical object data base\\[-1pt]
\noalign{\vspace{2pt}}
Other:\\
\quad \url{exoplanet.open.ac.uk}&			ExoPlanet news&				Monthly electronic newsletter\\
\noalign{\vspace{2pt}}
\hline
\end{tabular*}
\end{table*}

\subsection{Physical nature and definition}
As for Pluto's status in our own solar system, the precise definition of an exoplanet is not entirely trivial (and has been the subject of IAU recommendations and resolutions). Considerations involve the object's mass (whether below that of hydrogen/deuterium fusion), its formation mechanism (whether by disk accretion or protostellar gas cloud collapse), and details such as its shape and orbit. Broadly, a planet is considered to be the end product of disk accretion around a primary star or substar. Most discovered to date have masses below about 13~Jupiter masses (above which deuterium nuclear burning occurs).

\begin{enumerate}
\itemsep -3pt
\footnotesize
\setcounter{enumi}{\thetempref}

\item \label{item:whatisaplanet}
``Planetesimals to brown dwarfs: what is a planet?,'' 
G.~Basri and M.E.~Brown, Annual Review of Earth and Planetary Sciences {\bf 34}, 193--216 (2006). Wide-ranging discussion of the complex considerations involved in defining the term `planet'.~(I)

\setcounter{tempref}{\theenumi}
\end{enumerate} 

\section{Detection methods}

\subsection{Overview}
Direct {\it imaging\/} of exoplanets (the detection of a point-like image of the planet as a result of reflected star light) is highly challenging. For most of the exoplanets known today direct imaging is beyond current observational limits, due to a combination of the close planet--star proximity, and the intense `glare' of the host star swamping its reflected light. As a result, a number of other indirect methods have been developed and applied to their discovery and characterisation (Fig.~\ref{fig:detection-methods-montage}). 

As a planet orbits its host star, both planet and star orbit their common center of mass, leading to a periodic orbital motion of the star itself around the system's center of mass (the barycenter). Three `dynamical' detection methods exploit this motion (and consequently probe the exoplanet mass): measurement of the star's time-varying radial velocity, its position on the sky (astrometry) and, in a few special cases, timing signatures. Two other detection approaches are of central importance: photometric transits, and gravitational lensing.

\begin{figure}[t]
\centering
\includegraphics[width=1.0\linewidth]{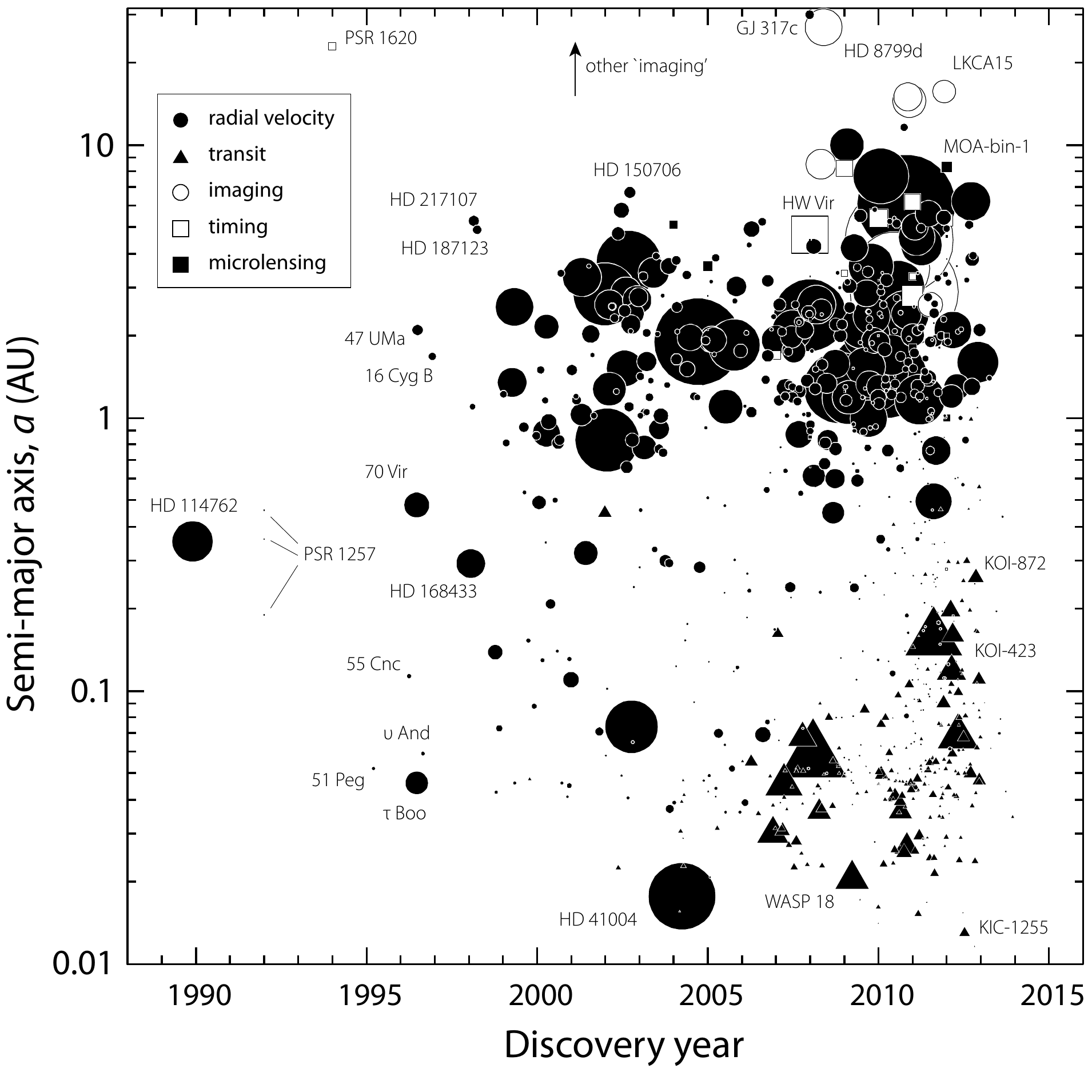}
\caption{Known exoplanets versus discovery date. The ordinate shows the semi-major axis in astronomical units (Earth's orbit defining 1\,AU). Symbols indicate detection method, with size proportional to exoplanet mass. The largest correspond to $\sim$10$M$(Jupiter), and the smallest to $\sim$1$M$(Earth), where $M$(Jupiter) $\sim300M$(Earth) $\sim0.001M$(Sun). A number of `imaging' exoplanets lie off the top of the plot, extending out to several hundred~AU (cf.\ Neptune's orbit of $a\sim30$\,AU).
\label{fig:discovery-year-versus-a}
}
\end{figure}

Fig.~\ref{fig:detection-methods} classifies these various methods in terms of exoplanet masses detectable, and the numbers discovered to date by each. Fig.~\ref{fig:discovery-year-versus-a} illustrates the discoveries over the past 20~years, as a function of exoplanet semi-major axis (ordinate), discovery method (symbol), and mass (symbol size). This diagram underlines a number of key points: 
(a)~radial velocity measurements led the initial discoveries in the early 1990s. They have grown in number, have progressively pushed to lower exoplanet masses, and have continued to dominate numerically at $a\gtrsim0.5$\,AU;
(b)~transit measurements were first made around 2000. Various surveys have led to increased detections over the past 10~years, and measurements from space have led to the method dominating at $a\lesssim0.3$\,AU;
(c)~other methods have resulted in smaller numbers of detections to date, but nevertheless probe crucial aspects, and are expected to grow in number and importance as new imaging initiatives on the ground, and astrometric measurements from space, develop over the coming years.

These specific detection methods are considered in further detail in the following sections.

 \subsection{Individual methods}

\subsubsection{Radial velocities}

The {\it radial velocity\/} detection method (Fig.~\ref{fig:detection-methods-montage}d) probes the detailed stellar motion as the result of one or more orbiting planets through the spectroscopic Doppler shifts of the star's spectral lines. The effect is present for all stars with orbiting planets unless the orbit is directly `face on' to us on the sky. For small planetary masses the perturbations are evidently smaller, and for wider orbits the temporal changes occur only over years or decades, both restricting the method's applicability according to the instrument's accuracy and stability. 
 
The current state-of-the-art reaches long-term accuracies of around 0.2\,ms$^{-1}$, sufficient to reach planetary masses of around 1~Earth mass. More than 500 exoplanets in 400 systems have been discovered up to late 2013 (of which some~100 have three or more planets). Around 20 radial velocity instruments across the world have contributed to the discoveries, with HARPS at the ESO 3.6-m telescope (ref.~\ref{item:harps}), and HIRES at the Keck 10-m telescope (ref.~\ref{item:hires}) dominating both in terms of numbers of discoveries and accuracy.
 
Using this method, exoplanets have been discovered around a large variety of stars, including main-sequence stars like the Sun, low-mass M~dwarfs, and metal-poor stars (including one of probable extragalactic origin). Exoplanets with unexpected properties have proven to be common. These include very closely orbiting `hot Jupiters' (which are considered unlikely to have formed {\it in situ}), giant planets several times the mass of Jupiter, orbits highly inclined to the host star's rotation axis, and even retrograde orbits. Multi-planet systems have provided a wealth of information on resonant systems, from which constraints on formation and migration processes have been inferred (see sect.~\ref{sec:formation}), as well as on dynamical architectures and orbital stability. Amongst the large number of resonant systems are the first, GJ~876, in which the two planets b~and~c are locked in a 2:1 mean motion resonance (see also ref.~\ref{item:gj876}), and the first Laplace resonance (three orbiting resonant planets), actually in the same system, in which planets c, b~and~e orbit in periods of 30.4, 61.1, and 126.6\,d (see also ref.~\ref{item:laplace}).
 
\begin{enumerate}
\itemsep -3pt
\footnotesize
\setcounter{enumi}{\thetempref}

\item
``A Jupiter-mass companion to a solar-type star,''
M.~Mayor, D.~Queloz, Nature {\bf 378}, 355--359 (1995). The first confirmed (Jupiter-mass) exoplanet around a main-sequence star, albeit in a remarkably short-period orbit of 4.2\,d (well inside Mercury's); this paper focused astronomers' attention on the prospects for exoplanet research.~(I)

\item \label{item:harps}
``Setting new standards with HARPS,''
M.~Mayor, F.~Pepe, D.~Queloz et al., The Messenger {\bf 114}, 20--24 (2003). Description of one of the two most prolific, and most accurate, of the radial velocity instruments.~(I/A)

\item \label{item:hires}
``HIRES: the high-resolution \'echelle spectrometer on the Keck 10-m telescope,''
S.S.~Vogt, S.L.~Allen, B.C~Bigelow et al., SPIE {\bf 2198}, 362--375 (1994). More technical details of the other prolific radial velocity instrument.~(I/A)

\item
``Five planets orbiting 55~Cnc,''
D.A.~Fischer, G.W.~Marcy, R.P.~Butler et al., ApJ {\bf 675}, 790--801 (2008). A system with five orbiting planets deduced from 18-yr of radial velocity measurements, including a discussion of how the signals of each are disentangled.~(I)

\item
``The HARPS search for southern extrasolar planets. XXVIII.~Up to seven planets orbiting HD~10180: probing the architecture of low-mass planetary systems,''
C.~Lovis, D.~S\'egransan, M.~Mayor et al., A\&A {\bf 528}, A112 (2011).
A system subsequently updated in 2012 to comprise nine orbiting planets, the most-populated known to date. General relativity effects and tidal dissipation are important in stabilising the innermost planet and the system as a whole.~(I)

\item
``The HARPS search for southern extrasolar planets. XXXIV.~Occurrence, mass distribution and orbital properties of super-Earths and Neptune-mass planets,''
M.~Mayor, M.~Marmier, C.~Lovis et al., A\&A, in press.
Illustrating the rate of advance, this recent paper announced 37~new exoplanets based on eight years of HARPS data.~(I)

\item
``An Earth-mass planet orbiting {$\alpha$}~Cen~B,''
X.~Dumusque, F.~Pepe, C.~Lovis et al., Nature {\bf 491}, 207--211 (2012).
Discovery of a planet around our nearest stellar system.
Later questioned by A.~Hatzes et al., ApJ {\bf 770}, 133 (2013), this illustrates 
the continuing importance of independent validation.~(I) 

\item 
Ref.~\ref{item:perryman}, Chapter~2 provides details of the principles, discoveries, and bibliography through to the end of 2010.~(I)

\item
\url{exoplanet.eu/catalog} and select `detected by radial velocity' to identify the exoplanets currently detected by this method, and the bibliography for each.
 
\setcounter{tempref}{\theenumi}
\end{enumerate} 

\subsubsection{Astrometry}

\begin{figure}[t]
\centering
\includegraphics[width=0.76\linewidth]{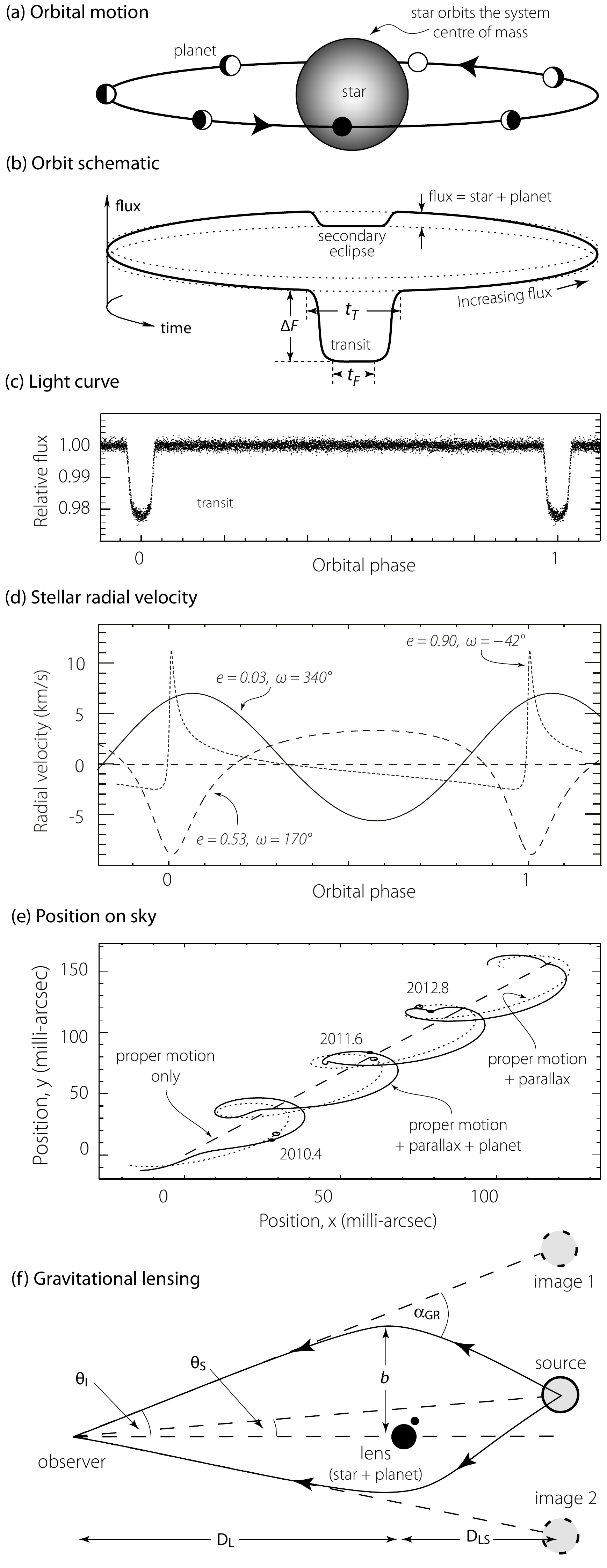}
\caption{Exoplanet detection principles (see text for details):
(a--c) transits and associated light curve;
(d)~the star's radial velocity as it orbits the system center-of-mass. Three different cases reflect different physical orbits (eccentricity, $e$), and viewing geometries (characterized by the `argument of pericenter', $\omega$);
(e)~schematic of how the star path on the sky can be revealed by astrometry;
(f)~gravitational lensing leading to light amplification of a background star.
\label{fig:detection-methods-montage}
}
\end{figure}

{\it Astrometric\/} detection (Fig.~\ref{fig:detection-methods-montage}e) exploits the host star's minuscule (side-to-side) positional oscillations on the sky as the result of one or more orbiting planets. The effect exists irrespective of the orbital projection geometry with respect to the observer. Current astrometric instruments are barely able to detect the expected positional motions, just one astrometric exoplanet discovery has been convincingly claimed to date (ref.~\ref{item:phases}), although observations from space (Hipparcos and Hubble Space Telescope) have made some preliminary advances. The situation will change substantially with ESA's Gaia space astrometry mission (launch due late 2013/early 2014).

\begin{enumerate}
\itemsep -3pt
\footnotesize
\setcounter{enumi}{\thetempref}

\item \label{item:phases}
``The PHASES differential astrometry data archive. V.~Candidate substellar companions to binary systems,''
M.W.~Muterspaugh, B.F.~Lane, S.R.~Kulkarni et al., AJ {\bf 140}, 1657--1671 (2010). The latest in a long history of claimed astrometric planet detections. All previous claims having been demonstrated to have been false, the authors state that {\it `these may represent either the first such companions detected, or the latest in the tragic history of this challenging approach.'} HD~176051\,b currently appears as the only astrometric discovery in \url{exoplanet.eu/catalog}.~(I)

\item
``New observational constraints on the $\upsilon$~And system with data from the Hubble Space Telescope and Hobby--Eberly Telescope,''
B.E.~McArthur, G.F.~Benedict, R.~Barnes et al., ApJ {\bf 715}, 1203--1220 (2010). Astrometry measurements at the sub-milliarcsec level are at the limits of present technology, but nevertheless illustrate the method's importance in establishing the degree of co-planarity of the orbits of multiple exoplanet systems, important for characterizing their formation and evolution.~(A)

\item
``Double-blind test program for astrometric planet detection with Gaia,''
S.~Casertano, M.G.~Lattanzi, A.~Sozzetti et al., A\&A {\bf 482}, 699--729 (2008). Estimates of the detectability of exoplanets with the Gaia space astrometry mission.~(I).

\item 
Ref.~\ref{item:perryman}, Chapter~3 provides details of the principles, discoveries, and bibliography through to the end of 2010.~(I)

\setcounter{tempref}{\theenumi}
\end{enumerate} 

\subsubsection{Timing}

In the {\it timing\/} method for exoplanet detection, any strictly periodic time signature of the host star will be perturbed by an orbiting planet, whose presence is revealed by nonperiodicity. Such phenomena are relatively uncommon, but include radial pulsations, pulsar-type radio emissions, or spectroscopic binary eclipses. 

The first reliable detection of planetary mass objects surrounding other stars were the three objects orbiting the pulsar PSR~B1257+12 reported in 1992. Known pulsar planets remain few in number (only one other has been discovered since), orbit stars at the end stages of their evolutionary lifetime, and may have been formed by very different mechanisms to those now known to surround main-sequence stars. The timing method has subsequently been applied to the detection of planets surrounding pulsating and eclipsing binary stars. 

\begin{enumerate}
\parsep 0pt\itemsep 0pt
\footnotesize
\setcounter{enumi}{\thetempref}

\item
``A planetary system around the millisecond pulsar PSR~B1257+12,''
A.~Wolszczan, D.A.~Frail,
Nature {\bf 355}, 145 (1992). The discovery of the first pulsar planets.~(I)

\item
``Discovery of pulsar planets,''
A.~Wolszczan,
New Astronomy Reviews {\bf 56}, 2 (2012). A recent perspective on the discovery by the same author.~(I)

\item 
``A giant planet orbiting the extreme horizontal branch star V391~Peg,''
R.~Silvotti, S.~Schuh, R.~Janulis et al., Nature {\bf 449}, 189--191 (2007). A late stellar evolutionary stage pulsating subdwarf, in which periodicities in the pulsation frequencies are considered evidence for the first exoplanet derived from pulsational timing irregularities.~(A)

\item
``The sdB+M eclipsing system HW~Vir and its circumbinary planets,''
J.W.~Lee, S.~Kim, C.~Kim et al., AJ {\bf 137}, 3181--3190 (2009). Confirmation of earlier hypotheses that the timing variations of this short period (2.8-hr) eclipsing binary are attributable to orbiting planets. Other planets orbiting eclipsing binary stars have since been reported.~(A)

\item 
Ref.~\ref{item:perryman}, Chapter~4 provides details of the principles, discoveries, and bibliography through to the end of 2010.~(I)

\item
\url{exoplanet.eu/catalog} and select `detected by timing' to identify the exoplanets currently detected by this method, and the bibliography for each.

\setcounter{tempref}{\theenumi}
\end{enumerate} 

\subsubsection{Microlensing}

Gravitational lensing, specifically in the domain referred to as {\it microlensing} in which distinct images are not resolved, provides an important if somewhat less intuitive method of exoplanet discovery (Fig.~\ref{fig:detection-methods-montage}f). Under conditions of essentially perfect alignment between a distant background star (providing an arbitrary source of light), an intermediate-distance foreground star with an orbiting planet, and the observer on Earth, the light from the background star is gravitationally `lensed' according to the prescriptions of general relativity, and significantly amplified both by the foreground star and its orbiting planet. Observers monitor large numbers of stars, and wait for serendipitous alignment conditions eventually brought about by the ever-changing space motions of the background star, foreground star, and Earth--Sun system. By recording the amplified light of the background star over the days or weeks of the evolving alignment, planets can be discovered, and their mass and orbital semi-major axis can be derived.  

\begin{enumerate}
\itemsep -3pt
\footnotesize
\setcounter{enumi}{\thetempref}

\item 
``The gravitational lens effect,''
S.~Refsdal, MNRAS {\bf 128}, 295--306 (1964). 
Gravitational lensing by a foreground object was first considered by Eddington (1920), Chwolson (1924), and Einstein (1936). The latter stated: {\it `Of course, there is no hope of observing this phenomenon directly'}. Refsdal gave formulae that have been applied and developed in various forms since.~(A)

\item
``Gravitational microlensing at large optical depth,''
B. Paczy{\'n}ski, ApJ {\bf 301}, 503--516 (1986). 
A paper that catalyzed microlensing studies of Galactic structure, and the associated search for exoplanets based on microlensing.~(A).

\item
``OGLE--2003--BLG--235/MOA--2003--BLG--53: a planetary microlensing event,''
I.A.~Bond, A.~Udalski, M.~Jaroszy{\'e}ski et al., ApJ {\bf 606}, L155--158 (2004). 
The first exoplanet detected by gravitational microlensing.~(A)

\item
``Discovery of a Jupiter/Saturn analogue with gravitational microlensing,''
B.S.~Gaudi, D.P.~Bennett, A.~Udalski et al., Science {\bf 319}, 927--930 (2008). 
The fifth exoplanet system detected by microlensing, and the first two-planet system. The complexity of the microlensing light curve even allowed orbital motions to be detected.~(A)

\item
``Frequency of solar-like systems and of ice and gas giants beyond the snow line from high-magnification microlensing events in 2005--2008,''
A.~Gould, S.~Dong, B.S.~Gaudi et al., ApJ {\bf 720},1073--1089 (2010). An example of important statistical inferences from the frequency of lensing events.~(I/A)

 \item
``Microlensing surveys for exoplanets,'' 
B.~S.~Gaudi, ARA\&A {\bf 50}, 411--453 (2012). 
Review of the concepts of microlensing planet searches and their practical application.~(I)

\item 
Ref.~\ref{item:perryman}, Chapter~5 provides details of the principles, discoveries, and bibliography through to the end of 2010.~(I)

\item
\url{exoplanet.eu/catalog} and select `detected by microlensing' to identify the exoplanets currently detected by this method, and the bibliography for each.

\setcounter{tempref}{\theenumi}
\end{enumerate} 

\subsubsection{Transits}

If the exoplanet orbit around its host star is fortuitously aligned such that its orbit plane is edge-on to our line-of-sight (Fig.~\ref{fig:detection-methods-montage}a--c), the planet may transit across the stellar disk, analogous to the transits of Mercury and Venus seen from Earth. Then its radius can be measured from the decrease in starlight, and its orbital period from the time between transits.

This method is of great importance to exoplanet studies since combining the radius of the planet with its mass (from dynamical studies), yields its density and hence (under certain assumptions) first indications of its composition. A number of other more subtle phenomena can be studied through the detailed form of the transit light curve: orbital eccentricity, planet surface gravity, star spots, the alignment of the orbit plane with the stellar rotation axis (via the Rossiter--McLaughlin effect, ref.~\ref{item:rmeffect}), planetary oblateness, apsidal precession, transit-timing variations due to other planets, and the possibility of detecting exoplanet satellites and rings. 

Of the dedicated ground-based surveys, HAT (Hungarian Automated Telescope, ref.~\ref{item:hat}) and WASP (Wide-Angle Search for Planets, ref.~\ref{item:wasp}) have together discovered some 150~exoplanets orbiting nearby bright stars, providing a wealth of physical insights due to the high signal-to-noise follow-up observations possible. From space, CoRoT (launched 2006, ref.~\ref{item:corot}) and in particular Kepler (launched 2009, ref.~\ref{item:kepler}) have together discovered many more. Although Kepler ceased nominal operations in May 2013, a substantial fraction of several thousand high-quality photometric candidates are likely to be considered as confirmed planets in due course. The Kepler mission is yielding unprecedented statistical insight into transiting systems (ref.~\ref{item:batalha}), and has discovered (amongst others) many multiple planet systems (frequently displaying orbital resonances), numerous `dynamically-packed' systems including one with six transiting planets (ref.~\ref{item:kepler11}), the first circumbinary planet (ref.~\ref{item:kepler16}), the first transiting planet smaller than Earth (ref.~\ref{item:kepler20}), and the first planet--planet eclipse (ref.~\ref{item:koi94}).

Spectroscopy around the time of the `secondary eclipse' (as the planet moves behind the star, Fig.~\ref{fig:detection-methods-montage}b) can reveal information about the planet's atmosphere, providing insight into its chemical composition, temperature, albedo, temperature inversions, and atmospheric circulation.

\begin{enumerate}
\itemsep -3pt
\footnotesize
\setcounter{enumi}{\thetempref}

\item
``Detection of planetary transits across a Sun-like star,''
D.~Charbonneau, T.M.~Brown, D.W.~Latham et al., ApJ {\bf 529}, L45--L48 (2000). 
The first exoplanet transit observed for a star (HD~209458) already identified as planet-hosting from earlier radial velocity measurement. Reported independently by G.W.~Henry et al., ApJ {\bf 529}, L41--L44 (2000).~(E)

\item \label{item:rmeffect}
``The long history of the Rossiter--McLaughlin effect and its recent applications,''
S.~Albrecht, IAU Symposium {\bf 282}, 379--384 (2012). An overview of the history of the effect, and its application to stellar and exoplanet observations.~(I)

\item \label{item:hat}
``HAT--P--34\,b -- HAT--P--37\,b: four transiting planets more massive than Jupiter orbiting moderately bright stars,''
G.A.~Bakos, J.D.~Hartman, G.~Torres et al., AJ {\bf 144}, 19--26 (2012). Some of the recent HAT discoveries, with a synopsis of the HAT transit survey.~(I)

\item \label{item:wasp}
``New transiting exoplanets from the SuperWASP-North survey,''
F.~Faedi, S.C.C.~Barros, D.~Pollacco et al., IAU Symposium {\bf 276}, 143--147 (2011). 
A review of the WASP--North survey, recent discoveries, and how they fit into the understanding of transiting exoplanet properties.~(I)

\item  \label{item:corot}
``Transiting exoplanets from the CoRoT space mission,''
C.~Cavarroc, C.~Moutou, D.~Gandolfi et al., Ap\&SS {\bf 337}, 511--529 (2012). 
A CoRoT satellite discovery paper describing candidate identification and confirmation.~(I)

\item \label{item:kepler}
``Kepler planet-detection mission: introduction and first results,''
W.J.~Borucki, D.~Koch, G.~Basri et al., Science {\bf 327}, 977--980 (2010). 
An overview of the mission.~(I)

\item \label{item:batalha}
``Planetary candidates observed by Kepler,''
N.M.~Batalha, J.F.~Rowe, S.T.~Bryson et al., ApJS {\bf 204}, 24 (2013).
Statistical results on the occurrence of planets and multiple systems from the first 16~months of Kepler data.~(I)

\item \label{item:kepler11}
``A closely packed system of low-mass, low-density planets transiting Kepler--11,''
J.J.~Lissauer, D.C.~Fabrycky, E.B.~Ford et al., Nature {\bf 470}, 53--58 (2011). One of the Kepler satellite transit discoveries with six~transiting planets (essentially coplanar) orbiting the same star. The inner five all lie within the orbital radius of Mercury.~(I)

\item \label{item:kepler16}
``Kepler--16: a transiting circumbinary planet,''
L.R.~Doyle, J.A.~Carter, D.C.~Fabrycky et al., Science {\bf 333}, 1602--1603 (2011). The first Kepler satellite transit discovery of a planet orbiting a binary star. Other circumbinary planets have since been discovered, and issues of their formation and orbit stability are active fields of investigation.~(I)

\item \label{item:kepler20}
``Two Earth-sized planets orbiting Kepler--20,''
F.~Fressin, G.~Torres, J.F.~Rowe et al., Nature {\bf 482}, 195--198 (2012). A demonstration that space transit observations can reach exoplanet masses as small as that of Earth.~(I)

\item \label{item:koi94}
``Planet--planet eclipse and the Rossiter--McLaughlin effect of a multiple transiting system,''
T.~Hirano, N.~Narita, B.~Sato et al., ApJ {\bf 759}, L36--L40 (2012). The Kepler light curve of this four-planet system captures a transit in which one of the exoplanets itself partially eclipses another.~(I)

\item
``Infrared transmission spectra for extrasolar giant planets,''
G.~Tinetti, M.-C.~Liang, A.~Vidal-Madjar et al., ApJ {\bf 654}, L99--L102 (2007). An introduction to the principles of transmission spectroscopy during an exoplanet's secondary eclipse, and early predictions of the observational capabilities of space instrumentation on Spitzer and JWST.~(I)

\item
``Transiting exoplanets with JWST,''
S.~Seager, D.~Deming, J.A.~Valenti, A\&SS Proceedings, 123--130 (2009). A review of the physical characteristics of transiting planets, results from Spitzer high-contrast exoplanet measurements, and examples of potential JWST observations.~(I)

\item 
Ref.~\ref{item:perryman}, Chapter~6 provides details of the principles, discoveries, and bibliography through to the end of 2010.~(I)

\item
\url{exoplanet.eu/catalog} and select `detected by transits' to identify the exoplanets currently detected by this method, and the bibliography for each.

\setcounter{tempref}{\theenumi}
\end{enumerate} 

\subsubsection{Imaging}

Imaging remains highly challenging because of the relative proximity between star and planet, and the enormous contrast ratio (a factor of $10^8$ or more) between the visible light from the star and that reflected from, or emitted by, the planet (the contrast ratio improves moving from the visible to the infrared spectral range). As a result only young, warm and self-luminous giant planets moving in a very wide (long-period) orbits have been detected to date. Specially-designed ground-based instruments (notably VLT--SPHERE and Gemini Planet Imager, ref.~\ref{item:gpi}) are expected to come on line in late-2013 to early-2014 to advance these observations. They are equipped with `extreme' adaptive optics (to minimize atmospheric scattering of the intense light from the host star) and coronagraphs (to attenuate the light from the star). 

Space missions, notably NASA's Terrestrial Planet Finder (TPF, ref.~\ref{item:tpf}) and ESA's Darwin (ref.~\ref{item:darwin}), have been studied over the past decade but are presently in hibernation due to their challenging technologies. Both coronagraphic and free-flying interferometric versions were considered, operating in either the visible or infrared (the choice being influenced by the spectral lines and intended atmospheric diagnostics being emphasized). 

\begin{enumerate}
\itemsep -3pt
\footnotesize
\setcounter{enumi}{\thetempref}

\item
``A probable giant planet imaged in the $\beta$~Pic disk: VLT--NACO deep L-band imaging,''
A.-M.~Lagrange, D.~Gratadour, G.~Chauvin et al., A\&A {\bf 493}, L21--L25 (2009).  Since the discovery of its dusty disk in 1984, $\beta$~Pic has become the prototype of young early-type planetary systems. Deep adaptive-optics L-band imaging using VLT--NaCo revealed one of the first exoplanet images: a faint point-like signal some 8\,AU from the star.~(I)

\item
``Direct imaging of multiple planets orbiting the star HR~8799,''
C.~Marois, B.~Macintosh, T.~Barman et al., Science {\bf 322}, 1348--1352 (2008). High-contrast observations with the Keck and Gemini telescopes revealed three planets orbiting the star HR~8799, with projected separations of 24, 38, and 68~AU, with multi-epoch data showing (counter-clockwise) orbital motion for all three imaged planets.~(I) 

\item \label{item:gpi}
``Experimental design for the Gemini Planet Imager,''
J.~McBride, J.R.~Graham, B.~Macintosh et al., PASP {\bf 123}, 692--708 (2011). The motivation, technology, and expected scientific harvest of this high-performance exoplanet imaging instrument, due to be commissioned during 2013.~(I)

\item \label{item:tpf}
``Prospects for Terrestrial Planet Finder,''
W.A.~Traub, S.~Shaklan, P.~Lawson, in {\bf In the Spirit of Bernard Lyot: The Direct Detection of Planets and Circumstellar Disks}, (University of California, Berkeley, 2007). One of the final presentations of NASA's TPF space imaging mission study in its coronagraphic and interferometric forms, now in hibernation.~(I)

\item \label{item:darwin}
``Darwin: an experimental astronomy mission to search for extrasolar planets,''
C.S.~Cockell, T.~Herbst, A.~L{\'e}ger et al., Experimental Astronomy {\bf 23}, 435--461 (2009). A summary of the scientific and design goals of the Darwin mission for exoplanet imaging from space, now in hibernation.~(I)

\item 
Ref.~\ref{item:perryman}, Chapter~7 provides details of the principles, discoveries, and bibliography through to the end of 2010.~(I)

\item
\url{exoplanet.eu/catalog} and select `detected by imaging' to identify the exoplanets currently detected by this method, and the bibliography for each.

\setcounter{tempref}{\theenumi}
\end{enumerate} 

It should be stressed that `imaging' of exoplanets implicitly refers to the detection of a point-like image of the planet. Resolved imaging of the planet surface is a substantially more challenging problem, presently considered as `beyond the state-of-the-art'. As described in ref.~\ref{item:woolf2001} {\it `the scientific benefit from this monstrously difficult task does not seem commensurate with the difficulty'}. Resolved imaging using `hypertelescopes' (ref.~\ref{item:labeyrie1996}) has also been considered.

\begin{enumerate}
\itemsep -3pt
\footnotesize
\setcounter{enumi}{\thetempref}

\item
``Multi-resolution element imaging of extrasolar Earth-like planets,''
P.L.~Bender, R.T.~Stebbins, J.~Geophys.\ Res.\ {\bf 101}, 9309--9312 (1996). Design of a separated spacecraft interferometer achieving visible light images with $10\times10$ resolution elements across an Earth-like planet at 10\,pc. The authors concluded that it would demand resources that would {\it `dwarf those of the Apollo Program or Space Station'}.~(I)

\item \label{item:woolf2001}
``Very large optics for the study of extrasolar terrestrial planets: Life Finder,''
N.J.~Woolf, Study Report, NASA Institute for Advanced Concepts (NASA, Atlanta, 2001).~(I)

\item \label{item:labeyrie1996}
``Resolved imaging of extrasolar planets with future 10--100~km optical interferometric arrays,''
A.~Labeyrie, A\&AS {\bf 118}, 517--524 (1996).
Argues that resolved images of exoplanets are in principle obtainable with ground-based arrays over 10~km baselines.~(A)

\setcounter{tempref}{\theenumi}
\end{enumerate} 

\section{Host stars}

Many detailed studies of the host stars of exoplanets have been made. These include their distances, luminosities, kinematic and dynamical properties derived from the Hipparcos space astrometry mission, abundance analyses to examine the occurrence of exoplanets as a function of metallicity and thereby probing their formation mechanism (ref.~\ref{item:haywood2009}), and specific studies of lithium and beryllium abundances to examine whether the surface layers of the host star has been `polluted' by the accretion of planetary matter (ref.~\ref{item:israelian2009}). Studies have examined the radio and X-ray emission properties of host stars with the aim of probing star--planet interactions and magnetospheric emission processes. Asteroseismology, the study of stellar oscillations, is being used to give independent constraints on fundamental stellar parameters such as mass, density, radius, age, rotation period, and chemical composition (ref.~\ref{item:christensen2010}).

The growing number of confirmed planets is leading to improved statistics on the frequency of planetary systems around stars of different types (refs~\ref{item:kopparapu2013}--\ref{item:howard2012}).
\begin{enumerate}
\itemsep -3pt
\footnotesize
\setcounter{enumi}{\thetempref}

\item \label{item:haywood2009}
``On the correlation between metallicity and the presence of giant planets,''
M.~Haywood, ApJ {\bf 698}, L1--L4 (2009). A summary of the dependency of the presence of giant planets on the metallicity of the host star, raising the possibility that these dependencies are related to the Galactic origin rather than the formation mechanism.~(I/A)

\item \label{item:israelian2009}
``Enhanced lithium depletion in Sun-like stars with orbiting planets,''
G.~Israelian, E.~Delgado Mena, N.C.~Santos et al., Nature {\bf 462}, 189--191 (2009). A study of the lithium abundance of planet hosting stars. Results suggest that the presence of planets may increase the amount of mixing and deepen the convective zone such that Li can be burned and therefore depleted.~(I/A)

\item \label{item:christensen2010}
``Asteroseismic investigation of known planet hosts in the Kepler field,''
J.~Christensen-Dalsgaard, H.~Kjeldsen, T.M.~Brown et al., ApJ {\bf 713}, L164--L168 (2010). An example of asteroseismology applied to exoplanet host stars, used to constrain fundamental host star properties, and hence determine better transiting planet radii by minimizing the uncertainties on stellar radii to which they are referred.~(I/A)

\item \label{item:kopparapu2013}
``A revised estimate of the occurrence rate of terrestrial planets in the habitable zones around Kepler M-dwarfs,''
R.K.~Kopparapu, ApJ {\bf 767}, 8 (2013).
A study combining Kepler detections with revised estimates of the `habitable zone' suggests that essentially one half of all (low-mass) M~dwarfs harbour Earth-sized planets within such an orbital region.~(I)

\item \label{item:fressin2013}
``The false positive rate of Kepler and the occurrence of planets,''
F.~Fressin, G.~Torres, D.~Charbonneau et al., ApJ {\bf 766}, 81 (2013).
The improving Kepler statistics suggests that nearly 20\% of main-sequence FGK stars have at least one Earth-sized planet with orbital periods up to 85~days, with no significant dependence on spectral type.~(I)

\item \label{item:howard2012}
``{Planet occurrence within 0.25~AU of solar-type stars from Kepler},''
A.W.~Howard, G.W.~Marcy, S.T.~Bryson et al., ApJS {\bf 201}, 15 (2012).
 A detailed discussion of planet occurrence around solar-type stars from the Kepler data.~(I)
 
\item 
Ref.~\ref{item:perryman}, Chapter~8 reviews the many detailed studies that have been carried out on exoplanet host stars through to the end of 2010.~(I)

\setcounter{tempref}{\theenumi}
\end{enumerate} 

\section{Brown dwarfs and free-floating planets}

A discussion of brown dwarfs must accompany any broad discussion of exoplanets for a number of reasons. Brown dwarfs are formed at the low-mass end of the star formation process, and studies of the environment of brown dwarfs therefore reveals further insight into the formation of disk and planetary systems around normal stars. Thus brown dwarfs are themselves now known to be accompanied by disks and probably their own (lower-mass) `planetary systems'. Understanding the formation and distribution of such brown dwarfs also helps to establish whether low-mass `free-floating' objects being discovered in star clusters are a byproduct of star formation, or whether they represent planets ejected from their own `solar system' as a result of dynamical evolutionary processes. An understanding of brown dwarfs also enters into the understanding of planet formation mechanisms within the protostellar disk. See also ref.~\ref{item:whatisaplanet}.

\begin{enumerate}
\itemsep -3pt
\footnotesize
\setcounter{enumi}{\thetempref}

\item
``The brown dwarf--exoplanet connection,''
I.N.~Reid and S.A.~Metchev, in {\bf Exoplanets: Detection, Formation, Properties, Habitability} (Springer--Praxis, Chichester, 2008), pp. 115--152. A review covering a historical introduction, the different techniques used to identify very low mass companions of stars, results from observational programs, and implications for brown dwarf and planetary formation mechanisms.~(I/A)

\item
``The discovery of brown dwarfs,''
G.~Basri, Scientific American {\bf 282}, 57--63 (2000). The challenging progress leading to the detection of the first brown dwarfs.~(E)

\item
``Observations of brown dwarfs,''
G.~Basri, ARA\&A {\bf 38}, 485--519 (2000). A detailed review of these objects which occupy the gap between the least massive star and the most massive planet.~(I/A)

\item 
Ref.~\ref{item:perryman}, Chapter~9 reviews the various relationships between exoplanets and brown dwarfs.~(I)

\setcounter{tempref}{\theenumi}
\end{enumerate} 

\section{Formation and evolution}
\label{sec:formation}

Treatments of exoplanet formation start with the complexities of star formation (ref.~\ref{item:starformation}), leading from protostars and protostellar collapse to disk formation in young stellar objects and the formation of protoplanetary disks. Within these, a succession of hierarchical processes leads in a `bottom-up' sequence to the formation of `terrestrial' mass planets: interstellar dust agglomerating into rocks (typical size $\sim1$\,m), thereafter into planetesimals ($\sim10^4-10^5$\,m), protoplanets ($\sim10^6$\,m), with more chaotic growth thereafter leading to objects of terrestrial planet size ($\sim10^7$\,m). There remains uncertainty whether the giant planets form by core accretion (rapid accretion of gas onto a massive core formed as for the terrestrial planets), or via gravitational disk instabilities, or some combination of the two. Thereafter, the gas disk from which the planets formed disperses. 

Either through viscous interaction with the residual gas disk, or through gravitational interaction with a residual planetesimal `sea', the planets so formed may then migrate inwards, or in some cases outwards, depending primarily on the object's mass. A further phase of planet--planet scattering may lead to qualitative changes in the architecture of the resulting system, through ejection of objects from the gravitational control of the host star, through rapid inward migration and perhaps infall onto the parent star, or through a variety of complex gravitational resonances (see also refs.~\ref{item:peale}--\ref{item:murray}) such as mean motion, spin--orbit, or Kozai resonances (as observed in our own solar system). Planets moving close-in to the host star are also subject to tidal evolution and tidal heating (ref.~\ref{item:ogilvie}). 

\begin{enumerate}
\itemsep -3pt
\footnotesize
\setcounter{enumi}{\thetempref}

\item \label{item:starformation}
``Theory of star formation,''
C.F.~McKee, E.C.~Ostriker, ARA\&A {\bf 45}, 565--687 (2007). A comprehensive review of the current understanding of star formation, outlining an overall theoretical and observational framework.~(I/A)

\item
{\bf Protoplanetary Dust: Astrophysical and Cosmochemical Perspectives},
D.A.~Apai, D.S.~Lauretta (Cambridge University Press, Cambridge, 2010). A comprehensive overview of the composition and evolutionary processes in the earliest stages of formation of protoplanetary structures.~(I/A)

\item
``Evolution of protoplanetary disk structures,''
F.J.~Ciesla, C.P.~Dullemond, in Protoplanetary Dust (Cambridge University Press, Cambridge, 2010), pp. 66--96. A review of protoplanetary disks, and how the gaseous and solid components evolve within them.~(I/A)

\item 
``The distribution of mass in the planetary system and solar nebula,''
S.J.~Weidenschilling, Ap\&SS {\bf 51}, 153--158 (1977). Early attempts to infer the structure of the solar nebula disk, taking account of the mass of material, and adding H/He at each location to produce a solar composition.~(I)

\item
``Dusty rings: signposts of recent planet formation''
S.J.~Kenyon, B.C.~Bromley, ApJ {\bf 577}, L35--L38 (2002). A contribution to the understanding that many older stars are accompanied by gas-poor debris disks arising from late-stage collisions of planetesimals.~(I/A)

\item \label{item:safronov}
{\bf Evolution of the Protoplanetary Cloud and Formation of the Earth and Planets},
V.S.~Safronov (Israel Program for Scientific Translations, Jerusalem, 1972). Many of the basic ideas central to the current picture of terrestrial planet formation were contained in this early monograph.~(I)

\item
``Gas disks to gas giants: simulating the birth of planetary systems,''
E.W.~Thommes, S.~Matsumura, F.A.~Rasio, Science, {\bf 321}, 814--817 (2008). Numerical simulations of the processes by which an initial protostellar disk converts itself into a small number of planetary bodies.~(I/A)

\item
{\bf Formation and Evolution of Exoplanets},
R.~Barnes, ed. (Wiley, Hoboken, 2010). An extensive review of the theory behind extrasolar planet formation. Individually-authored chapters cover different formation processes, planet--planet scattering, giant planets and brown dwarfs.~(I/A)

\item
``Giant planet formation by gravitational instability,''
A.P.~Boss, Science {\bf 276}, 1836--1839 (1997). Hydrodynamic calculations of protoplanetary disks showing that gravitational instability can rapidly form giant planets with modest cores of ice and rock.~(I)

\item
``Orbital migration and the frequency of giant planet formation,''
D.E.~Trilling, J.I.~Lunine, W.~Benz, A\&A {\bf 394}, 241--251 (2002).
A statistical study of the post-formation migration of giant planets, and the concepts underlying the late stages of assembly of planetary systems.~(I/A)

\item
``Planet--planet scattering leads to tightly packed planetary systems,''
S.N.~Raymond, R.~Barnes, D.~Veras et al., ApJ {\bf 696}, L98--L101 (2009). Planet--planet scattering during the final stages of planet formation can explain orbit eccentricities and their dynamical configurations.~(I/A)

\item
``Extrasolar planet population synthesis. Papers~I~and~II,''
C. Mordasini, Y. Alibert, W. Benz, A\&A {\bf 501}, 1139--1160 and  {\bf 501}, 1161--1184 (2009).  In `population synthesis' the statistical properties of the planetary population are used to constrain theoretical formation models.~(I/A)

\item \label{item:gj876}
``Dynamics and origin of the 2:1 orbital resonances of the GJ~876 planets,''
M.H.~Lee, S.J.~Peale, ApJ {\bf 567}, 596--609 (2002). Soon after the first discovered exoplanet orbit resonance, capture into resonance was explained by forced inward migration through disk viscosity.~(A)

\item \label{item:laplace}
``The Lick--Carnegie exoplanet survey: a Uranus-mass fourth planet for GJ~876 in an extrasolar Laplace configuration,''
E.J.~Rivera, G.~Laughlin, R.P.~Butler et al., ApJ {\bf 719}, 890--899 (2010). Of many known exoplanets in resonance orbits, this is the first Laplace resonance, in which three bodies have an integer ratio between their orbital periods.~(I)

\item 
``Transit timing variations in eccentric hierarchical triple exoplanetary systems,''
T.~Borkovits, S.~Csizmadia, E.~Forg{\'a}cs-Dajka et al., A\&A {\bf 528}, A53 (2011). Includes an accessible introduction to the phenomenon of Kozai cycles with tidal friction, one of the preferred theories for the formation of at least some of the observed hot Jupiters.~(A)

\item \label{item:ogilvie}
``Tidal dissipation in rotating giant planets,''
G.I.~Ogilvie, D.N.C.~Lin, ApJ {\bf 610}, 477--509 (2004).
While not the most recent study, this provides an accessible introduction to the effects of tidal interactions on orbital migration, resonant capture, and tidal heating.~(I/A)

\item 
Ref.~\ref{item:perryman}, Chapter~10 provides a broad review of all stages of exoplanet formation and evolution, while Section~2.6 covers the orbital resonances observed, how systems evolve into resonance, and the related topics of chaotic orbits and dynamical stability.~(I)

\setcounter{tempref}{\theenumi}
\end{enumerate} 

\section{Interiors and atmospheres}

A significant advance in the knowledge of exoplanet interiors and atmospheres has been made possible with the discovery of transiting exoplanets. Densities from their masses and radii are providing the first indications of their interior structure and composition, while broad-band brightness temperature measurements from transit photometry and spectroscopy are providing the first insights into their atmospheric composition and thermal transport processes. All of this has been substantially facilitated by the knowledge of interiors and irradiated atmospheres of solar system planets and satellites acquired over the last half century. 

Physical models of exoplanets span two extreme classes of object, and potentially much in between: the low-mass high-density `solid' planets dominated by metallic cores and silicate-rich and/or ice-rich mantles, and the high-mass low-density gas giants dominated by their massive accreted H/He envelopes. 

For the gas-rich giants, models of their interiors and models of their atmospheres are closely connected, and the most recent atmospheric models couple their emergent flux with their assembly by core accretion. Models of their interiors predict bulk properties such as the pressure-temperature relation, and their radii as a function of mass.  For close-in, highly irradiated gas giants, the additional external heat source has a significant effect on the pressure--temperature structure of the outer atmosphere. Combined with inferences on their probable bulk chemical composition, atmospheric models also predict broad-band colors and spectral features arising from specific atomic and molecular species.

Interior models of low-mass planets without massive gaseous envelopes aim to determine the mass--radius relation for a given internal composition, and the most likely internal composition given a specific planet's measured mass and radius. For terrestrial-type planets, a somewhat distinct problem is establishing the nature of their atmospheres: whether they might have acquired a gaseous envelope either by impact accretion or by outgassing, their likely composition, and whether such an atmosphere might have been retained or eroded over its evolutionary lifetime. 

\begin{enumerate}
\itemsep -3pt
\footnotesize
\setcounter{enumi}{\thetempref}

\item
``Exoplanet atmospheres,''
S.~Seager, D.~Deming, ARA\&A {\bf 48}, 631--672 (2010). Review of observations of various exoplanet atmospheres, describing the detection of molecular lines, observation of day/night temperature gradients, and constraints on vertical atmospheric structures.~(I)

\item
{\bf Exoplanet Atmospheres: Physical Processes},
S.~Seager (Princeton University Press, Princeton, 2010). Treats atmospheric composition and spectra, radiative transfer, and molecular and condensate opacities, emphasizing the major physical processes that govern planetary atmospheres.~(I/A)

\item
{\bf Principles of Planetary Climate},
R.T.~Pierrehumbert (Cambridge University Press, Cambridge, 2011). Principles relevant to understanding Earth's present/past climate, and solar system and extrasolar planets climates. Covers thermodynamics, infrared radiative transfer, scattering, and surface heat transfer.~(I/A)

\item
``Solar system abundances and condensation temperatures of the elements,''
K.~Lodders, ApJ {\bf 591}, 1220--1247 (2003). Thermodynamic modeling is used to predict the composition of the solar nebula, in elements and compounds (solids and gases), from initial elemental abundances.~(A)

\item
``The interiors of giant planets: models and outstanding questions,''
T.~Guillot, Annual Review of Earth and Planetary Sciences {\bf 33}, 493--530 (2005). Review of our knowledge of the solar system gas giants and inferences about the composition and structure of the giant exoplanets.~(I/A)

\item
``Planetary radii across five orders of magnitude in mass and stellar insolation: application to transits,''
J.J.~Fortney, M.S.~Marley, J.W.~Barnes, ApJ {\bf 659}, 1661--1672 (2007). Computation of model radii of H/He, water, rock, and iron planets, for masses from 0.01~Earth masses to 10 Jupiter masses at orbital distances of 0.02--10~AU.~(I/A)

\item
``Detailed models of super-Earths: how well can we infer bulk properties?,''
D.~Valencia, D.D.~Sasselov, R.J.~OÕConnell, ApJ {\bf 665}, 1413--1420 (2007). Interior models for terrestrial and ocean planets in the 1--10~Earth mass range.~(I/A)

\item
``Theoretical spectra and light curves of close-in extrasolar giant planets and comparison with data,''
A.~Burrows, J.~Budaj, I.~Hubeny, ApJ {\bf 678}, 1436--1457 (2008). Comprehensive atmospheric modeling program widely applied to brown dwarf and (irradiated) giant exoplanet atmospheres.~(A)

\item 
Ref.~\ref{item:perryman}, Chapter~11 provides a broad review of the various aspects of exoplanet interiors and atmospheres.~(I)

\setcounter{tempref}{\theenumi}
\end{enumerate} 

\null 
\section{Habitability}

Assessment of the suitability of a planet for supporting life is largely based on the knowledge of life on Earth. With the general consensus among biologists that carbon-based life requires water for its self-sustaining chemical reactions, the search for habitable planets has focused on identifying environments in which liquid water is likely to stable over billions of years. Even so, present discovery and characterisation methods yield essentially only the planet's equilibrium surface temperature. More complex atmospheric properties, for example, determine the fact that of the three planets in the Sun's `habitable zone', only one actually supports liquid surface water.

The new discipline of {\it astrobiology}, the study of the origin, evolution, distribution, and future of life in the Universe, encompasses the search for habitable environments in the solar system and beyond, and for biospheres that might be very different from Earth's. It is a cross-disciplinary effort, with knowledge being assimilated from astronomy, biology, chemistry, geography, geology, physics, and planetary science.

Many other conditions have been hypothesized as necessary for the development of life, largely based on the single life form known in the Universe, namely that on Earth. Hypothetical conditions rapidly become tied to the more philosophical discussions of the {\it anthropic principle}, which broadly states that the physical Universe must be compatible with conscious life that observes it.

The `search for extra-terrestrial intelligence' (SETI) is motivated by the belief that life, and intelligence, is likely to emerge under conditions resembling those on the early Earth. It is still generally perceived to lie somewhat at the edge of main-stream science -- perhaps like the search for exoplanets in the 1980s and early 1990s. 

\begin{enumerate}
\itemsep -3pt
\footnotesize
\setcounter{enumi}{\thetempref}

\item
``Habitable zones around main sequence stars,''
J.F.~Kasting, D.P.~Whitmire, R.T.~Reynolds, Icarus {\bf 101}, 108--128 (1993). One-dimensional climate modeling to estimate the habitable zone around our Sun and other main sequence stars.~(I) 

\item
``Terrestrial exoplanets: diversity, habitability and characterization,''
F.~Selsis, L.~Kaltenegger, J.~Paillet, Physica Scripta {\bf 130}, 014032 (2008). Future prospects for spectral characterization of habitable worlds.~(I/A)

\item
``Astrobiology: the study of the living universe,''
C.F.~Chyba, K.P.~Hand, ARA\&A {\bf 43}, 31--74 (2005). A review addressing the definition of life and the biological compatibility of the universe, examining the habitability of the Galaxy and of its constituent planets and moons.~(I)

\item
{\bf Astrobiology: A Multi-Disciplinary Approach},
J.I.~Lunine (Benjamin Cummings, San Francisco, 2005).
A thorough intermediate level summary of the various subdisciplines.~(I)

\item
{\bf Extrasolar Planets and Astrobiology},
C.A.~Scharf (University Science Books, Sausalito, 2009). Summarizes star and planet formation, and exoplanet detection, thereafter focusing on the development and search for life, and detailed considerations of the habitable zone.~(I)

\item
``The anthropic principle,''
Y.V.~Balashov, American Journal of Physics {\bf 59}, 1069--1076 (1991). Although published more than 20~years ago, this remains an excellent introduction to this field of somewhat philosophical enquiry.~(I)

\item
``Life, the Universe, and SETI in a nutshell,''
J.C.~Tarter, in `Bioastronomy 2002: Life Among the Stars', IAU Symposium {\bf 213}, 397--407 (2004). An overview of the state of SETI programs worldwide.~(E/I)

\item
``SETI: the early days and now,''
F.D.~Drake, in {\bf Frontiers of Astrophysics: A Celebration of NRAOÕs 50th Anniversary}, ASP Conf Ser {\bf 395}, 213--224 (2008). Coverage of 50~years of SETI searches.~(E)

\setcounter{tempref}{\theenumi}
\end{enumerate} 

\null
\section{The solar system}

Theories of exoplanet formation can be confronted with a wealth of observational constraints from the solar system. Conversely, the insights gained from exoplanet modeling have thrown considerable light on the origin and structure of the solar system. Amongst others, advances have been made in our understanding of the formation of the Moon (ref.~\ref{item:canup}), impact cratering (ref.~\ref{item:gomes}), orbital resonances (refs.~\ref{item:peale}--\ref{item:murray}), the origin of water on Earth (ref.~\ref{item:morbidelli}), the arrangement of planetary and solar obliquities (ref.~\ref{item:tremaine}), and planetesimal migration (ref.~\ref{item:tsiganis}). See also ref.~\ref{item:safronov} for an overview of the formation of the Earth and planets according to the `solar nebula' theory.

\begin{enumerate}
\itemsep -3pt
\footnotesize
\setcounter{enumi}{\thetempref}

\item \label{item:canup}
``Forming a moon with an Earth-like composition via a giant impact,''
R.M.~Canup, Science {\bf 338}, 1052--1055 (2012). Details of the current paradigm that the Moon formed from an oblique and late-stage giant collision between the Earth and a Mars-mass protoplanet.~(I)

\item \label{item:gomes}
``Origin of the cataclysmic Late Heavy Bombardment period of the terrestrial planets,''
R.S.~Gomes, H.F.~Levison, K.~Tsiganis et al., Nature {\bf 435}, 466--469 (2005). Associates a hypothesized migrating Saturn to the `late heavy bombardment' (some 4\,Gyr ago) of the Moon, and by inference to the cratering of Mercury, Venus, Earth, and Mars.~(I)

\item \label{item:peale}
``Orbital resonances in the solar system,''
S.J.~Peale, ARA\&A {\bf 14}, 215--246 (1976). An illumintaing account of resonance phenomena, also relevant for exoplanets.~(I/A)

\item \label{item:murray}
{\bf Solar System Dynamics},
C.D.~Murray, S.F.~Dermott (Cambridge University Press, Cambridge, 2000). An extensive mathematical treatment of the dynamical features of the solar system, similarly relevant for exoplanet orbits.~(A)

\item \label{item:morbidelli}
``Source regions and time scales for the delivery of water to Earth,''
A.~Morbidelli, J.~Chambers, J.~Lunine et al., Meteoritics \& Planetary Science {\bf 35}, 1309--1320 (2000). The authors argue that asteroids and comets from the Jupiter--Saturn region were the first deliverers to Earth, with the bulk subsequently carried by a few planetary embryos.~(I/A) 

\item \label{item:tremaine}
``On the origin of the obliquities of the outer planets,''
S.~Tremaine, Icarus {\bf 89}, 85--92 (1991). Links nonzero obliquities to asymmetric infall during protostellar collapse.~(I)

\item \label{item:tsiganis}
``Origin of the orbital architecture of the giant planets of the solar system,''
K.~Tsiganis, R.~Gomes, A. Morbidelli et al., Nature {\bf 435}, 459--461 (2005). This `Nice' model reproduces key characteristics of the giant planet orbits as a result of a Jupiter--Saturn resonance crossing as the giant planets migrated through interaction with a planetesimal disk.~(I)

\item
``Existence of collisional trajectories of Mercury, Mars and Venus with the Earth,''
J.~Laskar, M.~Gastineau, Nature {\bf 459}, 817--819 (2009). Long-term numerical integration that shows how the solar system's long-term stability depends sensitively on initial conditions.~(I)

\item
{Emergence of a habitable planet}
K. J. Zahnle, N. Arndt, C. Cockell et al., Space Sci.\ Rev.\ {\bf 129}, 35--78 (2007). A possible timeline of Earth's early atmosphere, starting with the Moon-forming impact, and extending over the 0.8\,Gyr of the Hadean (or pre-Archean) eon.~(I)

\item 
Ref.~\ref{item:perryman}, Chapter~12 provides more detail and literature references on these various aspects.~(I)

\setcounter{tempref}{\theenumi}
\end{enumerate} 

\section*{Acronyms}
Acronyms for organizations and instruments used in this article include:\\
CoRoT: Convection Rotation and planetary Transits\\
ESO: European Southern Observatory\\
HARPS: High Accuracy Radial Velocity Planet Searcher\\
HAT: Hungarian Automated Telescope\\
HIRES: High-Resolution Spectrograph (Keck)\\
IAU: International Astronomical Union\\
OGLE: Optical Gravitational Lensing Experiment\\
SPHERE: Spectro-Polarimetric High-contrast Exoplanet \phantom{xxxxxxxx} Research\\
VLT: Very Large Telescope (ESO)\\
WASP: Wide Angle Search for Planets

\vspace{10pt}
The potentially perplexing range of star and planet names draws on a range of etymologies, based on constellation (e.g.\ $\beta$~Pic, HW~Vir), major star catalogues including GJ (Gliese--Jahreiss), HD (Henry Draper), HIP (Hipparcos), and PSR (pulsar), or on planet detection instruments, either on ground (such as HAT, OGLE, WASP) or in space (notably CoRoT, Kepler).

\vspace{-10pt}
\begin{acknowledgments}

This article was completed during a Bohdan Paczy\'nski Visitorship at the Department of Astrophysical Sciences, Princeton. I am particularly grateful to David Spergel, Michael Strauss and Robert Lupton for this invitation. I thank the six anonymous referees for their valuable comments on the draft manuscript. 

\end{acknowledgments}

{
\bibliography{../../CUP-Exoplanet-Review-2011/biball-to-2010,../../Exoplanet-Handbook-second-edition/oct2010-dec2012} 
}

\end{document}